\documentclass[submission,copyright,creativecommons]{eptcs}
\providecommand{\event}{LFMTP 2021} 
\usepackage{underscore}           

\usepackage{proof}
\usepackage{amssymb}
\usepackage{stmaryrd}
\usepackage{amsmath}

\newcommand\rght[1]{\lceil #1 \rceil}
\newcommand\lft[1]{\lfloor #1 \rfloor}

\title{Facilitating Meta-Theory Reasoning\\
  (Invited Paper)}

\author{
  Giselle Reis
  \institute{Carnegie Mellon University in Qatar}
  \email{giselle@cmu.edu}
}

\def\titlerunning{Facilitating Meta-Theory Reasoning}
\def\authorrunning{Giselle Reis}

\begin{document}
\maketitle

\begin{abstract}
Structural proof theory is praised for being a symbolic approach to
reasoning and proofs, in which one can define schemas for reasoning
steps and manipulate proofs as a mathematical structure.
For this to be possible, proof systems must be designed as a set of
rules such that proofs using those rules are correct \emph{by
construction}. 
Therefore, one must consider all ways these rules can interact and
prove that they satisfy certain properties which makes them
``well-behaved''.
This is called the \emph{meta-theory} of a proof system.

Meta-theory proofs typically involve many cases on structures with
lots of symbols.
The majority of cases are usually quite similar, and when a proof
fails, it might be because of a sub-case on a very specific
configuration of rules.
Developing these proofs by hand is tedious and error-prone, and their
combinatorial nature suggests they could be automated.

There are various approaches on how to automate, either partially or
completely, meta-theory proofs.
In this paper, I will present some techniques that I have been involved
in for facilitating meta-theory reasoning.
\end{abstract}

\section{Introduction}

%
Structural proof theory is a branch of logic that studies proofs as
mathematical objects, understanding the kinds of operations and
transformations that can be done with them.
To do that, one must represent proofs using a regular and unambiguous
structure, which is constructed from a fixed set of rules.
This set of rules is called a \emph{proof calculus}, and there are
many different ones.
One of the most popular calculi used nowadays is \emph{sequent
calculus}~\cite{gentzen35} (and its variations).

%
In its simplest form, a sequent is written $\Gamma \vdash \Delta$,
where $\Gamma$ and $\Delta$ are sets or multisets of formulas
(depending on the logic).
The interpretation of a sequent is that the conjunction of formulas in
$\Gamma$ implies the disjunction of formulas in $\Delta$.
Rules in a sequent calculus are written:
\[
\infer[\mathsf{name}]{C}{P_1 & \dots & P_n}
\]

\noindent
where $C$ is the \emph{conclusion} sequent, and $P_i$ are the
\emph{premise} sequents.
When there are no premises, the rule is called an \emph{axiom}.
Sequent calculus proofs are trees where each node is an instance of a
rule.
A set of rules is considered a ``good'' or ``well-behaved'' sequent
calculus system if it satisfies a few properties.
Among the most important ones are:
\begin{enumerate} 
  \item identity expansion: the system is able to prove $A \vdash A$
  for any formula $A$; and 

  \item cut elimination: if $\Gamma \vdash \Delta, A$ and $\Gamma, A
  \vdash \Delta$, then $\Gamma \vdash \Delta$.
\end{enumerate}
A corollary of cut elimination is the calculus' \emph{consistency},
i.e. it cannot derive false.
But there are also other properties which are interesting to show,
such as rule invertibility and permutability.
All these properties are called the \emph{meta-theory} of a proof
system, and they can be used as lemmas in each others' proof, or
justifications for sound proof transformations or proof search
optimizations.
Meta-theory proofs are typically done via structural induction on
proof trees, formulas, or both.
The number of cases is combinatorial on the number of rules and/or
connectives, but cases tend to follow the same argument, with only a
few more involved ones.

%
The appeal of sequent calculus is its versatility and uniformity:
there are sequent calculi for a great number of logics, and they are
formed by rules which are usually very similar.
This similarity is very convenient, specially when it comes to
meta-theory proofs.
Since different logics share the same rules, parts of proofs can be
reused from one system to the other.
At first, meta-theory proofs are mostly developed by hand.
After seeing the same cases over and over again, we become more and
more confident that they will work out\footnote{If you have read
Kahneman's book, system 1 takes over.} and skip more and more steps.
Coupled with the sheer complexity and size of such proofs, we end up
missing cases and making mistakes, which need to be corrected later.

%
%
%
A proof of cut-elimination for full intuitionistic linear logic (FILL) was shown
to have a mistake in~\cite{Bierman96}, and the authors of the proof have later
published a full corrected version~\cite{Brauener96}.
A proof of cut-elimination for the sequent calculus GLS$_V$ for the provability
logic GL was the source of much controversy until this was resolved
in~\cite{Gore12} and formalized in~\cite{Dawson10} using Isabelle/HOL.
More recently, another proof of cut elimination for the provability
logic GLS was proposed~\cite{brighton15}, but the inductive measures
used were not appropriate.
Upon formalizing the proof~\cite{gore21}, other researchers not only
found the mistake, but were able to get a more self-contained proof.
Several sequent calculi proposed for bi-intuitionistic logic were ``proved'' to
enjoy cut-elimination when, in fact, they did not. The mistake is analysed and
fixed in~\cite{Pinto09}.
An error in the cut-elimination proof for modal logic nested systems
was corrected in~\cite{Marin14}.

%
This situation has led many researchers to look for easier and less
error prone methods for proving cut elimination.
In this paper, I am going to discuss three different approaches:
logical frameworks, formalization, and user-friendly implementations.
I will focus on those developments where I have been involved in,
but I will also mention relevant (though non-exhaustive) related work.

\section{Logical Frameworks}

In its most general form, a logical framework can be defined as a
specification language with a reasoning engine, which is capable of
reasoning about specifications written in the language.
Formalizing meta-theory proofs in a logical framework involves writing
the proof system in the specification language, and expressing
properties such as cut elimination in a sentence that can be decided
by the reasoning engine.

%
A classic example is the proof of cut elimination for the
intuitionistic sequent calculus LJ in the logical framework
LF~\cite{lf93}.
LF's specification language is a type theory, so LJ is specified by
writing each of its rules as an appropriate type.
LJ's cut elimination proof, in its turn, is written by using another
type for each proof case.
LF's reasoning engine is thus able to infer coverage and termination
of the proof.
If one trusts LF's checker, then one can be sure that the specified
cut elimination proof holds.
Unfortunately, LF's method does not translate so elegantly to other
logics, and fitting a sequent calculus system and its cut elimination
proof in LF's type theory can be quite an involved exercise.

%
The L-framework uses an implementation of rewriting logic (Maude) to
check meta-theory properties of sequent calculus
systems~\cite{OlarteRLA18}.
Each inference rule is specified as a rewriting rule, and
meta-properties are represented as reachability goals.
Different meta-properties are parametrized by different rewriting
rules.
These include a number of rewriting rules representing the proof
transformations relevant to that meta-property, and also the rewriting
rules corresponding to the sequent calculus system.
The rewriting logic implementation is responsible for solving the
reachability goal, and thus establishing whether the meta-property
holds.
This technique is sound but not complete.
If the meta-theory proof follows a more exotic strategy, then it is
likely that the pre-determined set of transformations in the
L-framework will not be enough to decide reachability.
In spite of this (expected) limitation, the L-framework was used to
show a number of meta-properties (invertibility, permutability,
cut-elimination, etc) of various sequent calculus systems, including
single and multi conclusion intuitionistic logic, classical logic,
linear logic, and modal logics.

%
Linear logic was proposed as a framework for reasoning about
sequent calculus systems in~\cite{llinda}. 
In this technique, each rule in sequent calculus is encoded as a
linear logic formula.
%
%
Meta-properties such as cut-elimination and identity expansion are
equivalent to (decidable) properties of the encoding.
Therefore, given a set of linear logic formulas representing the
encoding of a sequent calculus system, cut elimination can be decided
using bounded proof search in linear logic.
This method was used to prove meta-properties of linear, classical,
and intuitionistic logics.
It turns out that the linear logic framework cannot capture so easily
sequent calculus systems with rules that have side conditions in the
context, such as:
\[
\infer[\square_R]
  {\square \Gamma, \Gamma' \vdash \square A, \Delta}
  {\square \Gamma \vdash A}
\]
To be able to specify rules like this, we can use \emph{subexponential
linear logic} (SELL).

\subsection{(Subexponential) Linear Logic}

Linear logic is a refinement of classical logic in the sense that it
allows a finer control over structural rules.
In this logic, formulas that can be contracted or weakened (on the
right side) are marked with the exponential operator $?$, called
question mark.
The dual of $?$ is $!$, called bang.
As a result, there are two kinds of conjunction and disjunction: an
additive and a multiplicative, which differ on how the context is
split between premises.
The one-sided calculus for classical linear logic is depicted in
Figure~\ref{fig:ll}.
The context is formed by $\Gamma$, containing formulas that can be
contracted and weakened (i.e. those that were under $?$), and
$\Delta$, containing the other formulas.
A relevant rule for what follows is $!$, also called \emph{promotion}.
Observe that a formula under $!$ can only be used if no other formulas
exist in $\Delta$.

\begin{figure}
{\small
\[
\infer[\binampersand]{\vdash \Gamma ; \Delta, A \mathbin\binampersand B}{
  \vdash \Gamma ; \Delta, A
  &
  \vdash \Gamma ; \Delta, B
}
\qquad
\infer[\bindnasrepma]{\vdash \Gamma ; \Delta, A \mathbin\bindnasrepma B}{
  \vdash \Gamma ; \Delta, A, B
}
\qquad
\infer[\top]{\vdash \Gamma ; \Delta, \top}{}
\qquad
\infer[\bot]{\vdash \Gamma ; \Delta, \bot}{\vdash \Gamma ; \Delta}
\]
\[
\infer[\otimes]{\vdash \Gamma ; \Delta_1, \Delta_2, A \otimes B}{
  \vdash \Gamma ; \Delta_1, A
  &
  \vdash \Gamma ; \Delta_2, B
}
\qquad
\infer[\oplus_i]{\vdash \Gamma ; \Delta, A_1 \oplus A_2}{
  \vdash \Gamma ; \Delta, A_i
}
\qquad
\infer[1]{\vdash \Gamma ; 1}{}
\qquad
\infer[\mathsf{init}]{\vdash \Gamma ; a, a^\bot}{}
\]
\[
\infer[?]{\vdash \Gamma ; \Delta, ?A}{\vdash \Gamma, A ; \Delta}
\qquad
\infer[!]{\vdash \Gamma ; !A}{\vdash \Gamma ; A}
\qquad
\infer[\mathsf{copy}]{\vdash \Gamma, A ; \Delta}{\vdash \Gamma, A ; \Delta, A}
\]
}
\caption{One-sided dyadic sequent calculus for classical linear logic.}
\label{fig:ll}
\end{figure}

Subexponential linear logic (SELL) extends linear logic by allowing
multiple \emph{indexed} exponential operators $!^a,
?^a$~\cite{nigam09phd}.
Each index may or may not allow the rules of contraction and
weakening, and they are organized in a pre-order $\preceq$.
Since there are formulas under different $?^a$, the sequent in SELL is
composed of several contexts, one for each subexponential index.
Assuming subexponential indices $1$ to $n$, the rules involving
subexponentials are modified as follows.
Rule $?$ stores the formula in the appropriate context: 
\[
\infer[?]{\vdash \Gamma_1 ; ... ; \Gamma_i ; ... ; \Gamma_n ; \Delta, ?^i A}{
  \vdash \Gamma_1 ; ... ; \Gamma_i, A ; ... ; \Gamma_n ; \Delta
}
\]

Formulas under $!^i$ can only be used if all contexts corresponding to
indices $k$ such that $i \npreceq k$ and $\Delta$ are empty:
\[
\infer[! \text{ if } \Gamma_k = \emptyset \text{ for } i \npreceq k]{\vdash \Gamma_1 ; ... ; \Gamma_n ; !^i A}{
  \vdash \Gamma_1 ; ... ; \Gamma_n ; A
}
\]

Formulas in $\Gamma_i$ will only retain a copy in the context if $i$
is an index that allows contraction:
\[
\infer[\mathsf{copy} \text{ if } i \text{ allows contraction}]
  {\vdash \Gamma_1 ; ... ; \Gamma_i, A ; ... ; \Gamma_n ; \Delta}
  {\vdash \Gamma_1 ; ... ; \Gamma_i, A ; ... ; \Gamma_n ; \Delta, A}
\]
\[
\infer[\mathsf{use} \text{ if } i \text{ does not allow contraction}]
  {\vdash \Gamma_1 ; ... ; \Gamma_i, A ; ... ; \Gamma_n ; \Delta}
  {\vdash \Gamma_1 ; ... ; \Gamma_i ; ... ; \Gamma_n ; \Delta, A}
\]

\subsection{Encoding}

Using the new promotion rule, and the flexibility in choosing the
subexponential indices and their properties, we can encode rules such
as:
\[
\infer[\square_R]
  {\square \Gamma, \Gamma' \vdash \square A, \Delta}
  {\square \Gamma \vdash A}
\]

The encoding in linear logic uses two predicate symbols that map
object logic formulas into linear logic predicates: $\lft{\cdot}$ and
$\rght{\cdot}$.
$\lft{A}$ indicates that $A$ is an object logic formula that is on the
left side of the object logic sequent.
$\rght{A}$ indicates that $A$ is an object logic formula that is on
the right side of the object logic sequent.
Therefore, the sequent $A_1, ..., A_n \vdash B_1, ..., B_m$ in the
object logic is encoded, roughly, as the linear logic sequent:
$\vdash \lft{A_1}, ..., \lft{A_n}, \rght{B_1}, ..., \rght{B_m}$.
In addition to these formulas, the linear logic sequent also includes
a set $\mathcal{T}$ of LL formulas that encode the sequent calculus rules
of the considered logic.
The derivation of a formula in $\mathcal{T}$ mimics the application of
the corresponding rule in the object logic.
This general scheme is depicted in Figure~\ref{fig:encoding-scheme}.

\begin{figure}
\[
\infer[\wedge_R]{\Gamma \vdash A \wedge B}{
  \Gamma \vdash A & \Gamma \vdash B
}
\quad
\rightsquigarrow
\quad
\deduce{\vdash \mathcal{T}, \lft{\Gamma}, \rght{A \wedge B}}{
  \deduce{\vdots}{
    \vdash \mathcal{T}, \lft{\Gamma}, \rght{A}
    &
    \vdash \mathcal{T}, \lft{\Gamma}, \rght{B}
  }
}
\]
\caption{The encoding of the object logic rule $\wedge_R$ is a linear
logic formula whose derivation results on the right side tree.}
\label{fig:encoding-scheme}
\end{figure}

The main idea behind the encoding is to choose indices such that the
structure of the SELL sequent 
$\vdash \Gamma_1 ; ... ; \Gamma_n ; \Delta$ 
contains the parts of the context that need to be distinguished in the
object logic.
For the example above, we have a subexponential index to store boxed
formulas on the left (called $\square_l$), one to store all other
formulas on the left (called $l$), and one to store all formulas on
the right (called $r$)~\cite{NigamJLC16}.
Using the appropriate pre-order relation between those, we can force
the deletion of formulas in $l$ and $r$, when deriving the formula
corresponding to the encoding of $\square_R$.

The rule $\square_R$ above is encoded as the following SELL formula
(which resembles a Horn clause):
\[
\rght{\square A}^\bot \otimes !^{\square_l} ?^r \rght{A}
\]
If we use a pre-order where indices $\square_l$, $l$, and $r$ are not
related and allow contraction and weakening, the derivation of the
formula above in SELL is:
\[
\infer[\otimes]{\vdash \lft{\Gamma_l} ; 
                \lft{\Gamma_{\square_l}} ;
                \rght{\Delta}, \rght{\square A} ; \rght{\square A}^\bot \otimes !^{\square_l} ?^r \rght{A}}
{
  \infer[\mathsf{copy}]{\vdash \lft{\Gamma_l} ; \lft{\Gamma_{\square_l}} ; 
                               \rght{\Delta}, \rght{\square A} ; \rght{\square A}^\bot}
  {
    \infer[\mathsf{init}]{\vdash \lft{\Gamma_l} ; \lft{\Gamma_{\square_l}} ; 
                               \rght{\Delta}, \rght{\square A} ;
                               \rght{\square A}, \rght{\square A}^\bot}{}
  }
  &
  \infer[!]{\vdash \lft{\Gamma_l} ; \lft{\Gamma_{\square_l}} ; 
                  \rght{\Delta}, \rght{\square A} ; !^{\square_l} ?^r \rght{A}}
  {
    \infer[?]{\vdash \cdot ; \lft{\Gamma_{\square_l}} ; \cdot ; ?^r \rght{A}}
    {
      \vdash \cdot ; \lft{\Gamma_{\square_l}} ; \rght{A} ; \cdot
    }
  }
}
\]

Observe how the application of rule $!$ removes precisely the formulas
that are weakened in the $\square_R$ rule, and how the open premise
corresponds to the premise of the rule.

\subsection{Meta-theory}

Given a set of linear logic formulas encoding rules of a sequent
calculus system, \cite{llinda}~defined decidable conditions on the
formulas that translate into the meta-properties identity expansion
and cut elimination.
The same criteria could be used for identity expansion of sequent
calculus systems encoded in SELL, but not for cut elimination.

Showing cut elimination of systems encoded in LL is split into two
parts:
\begin{enumerate}
  \item\label{reduce} show that cuts can be reduced to atomic cuts;
  \item\label{remove} show that atomic cuts can be eliminated.
\end{enumerate}

Step~\ref{reduce} relies on the fact that specifications of dual
rules are dual LL formulas, and on cut elimination in LL itself.
Step~\ref{remove} is shown by mimicking Gentzen's reduction rules
that permute atomic cuts until $\mathsf{init}$ rules, and then remove
the cuts.

Cut is encoded as the linear logic formula $\rght{A} \otimes \lft{A}$,
so its permutation in the object logic is equivalent to he permutation
of this formula's derivation in LL.
It is shown in~\cite{llinda} that this permutation is possible.

In the case of systems encoded in SELL, the general shape of the cut
rule is:
\[
!^a ?^b \rght{A} \otimes !^c ?^d \lft{A}
\]

\noindent
where $!^a$ and $!^c$ may or may not occur.

The presence of subexponentials is necessary, since sequent calculus
systems with more complicated contexts may have cut rules that impose
restrictions on the context, or may need more than one cut rule.
As a result, the cut elimination argument for those systems may be
more involved.
They may require proof transformations to be done on a specific order
(e.g. permute the cut on the left branch before the right), or may
involve other transformations (e.g. permuting rules down the proof
instead of permuting the cut up), or may use more complicated
induction measures (e.g. a distinction between a ``light'' and
``heavy'' cut).

It is thus not a surprise that the elegant cut elimination criteria
from~\cite{llinda} does not translate so nicely to SELL encodings.
The presence of subexponentials on the formula corresponding to the
cut rule may prevent the derivation of this formula from permuting, so
the steps that require permutation of cut need to be looked at
carefully.
Therefore, showing cut elimination of systems encoded in SELL is split
into three parts:

\begin{enumerate}
  \item\label{principal} show that the cut can become \emph{principal}
  (i.e. the rules immediately above the cut operate on the cut
  formula);
  \item\label{atomic} show that principal cuts can become atomic cuts;
  \item\label{nocuts} show that atomic cuts can be eliminated.
\end{enumerate}

Step~\ref{atomic} is shown as in~\cite{llinda}, relying on the duality
of formulas for dual rules and cut elimination in SELL.
We have identified simple conditions for when step~\ref{nocuts} can be
performed, based on which subexponentials indices occur on the
encoding of the rules and how they are related in the pre-order.
However, step~\ref{principal} turned out to be quite complicated.
The reason was already alluded to before: making cuts principal may
involve permutations of the cut rule using particular strategies,
permutations of other rules, or transformations of one cut into
another.
We have identified conditions that allow for some of these
transformations in~\cite{NigamJLC16}, but it is unlikely that a
general criteria that encompasses this variety of operations
exists.

\begin{paragraph}{Permutation lemmas}
Step~\ref{principal} above may involve a number of permutation lemmas
between rules, so we focused our efforts in finding an automated
method to check for these transformations.
Using answer set programming (ASP) and the encoding of rules in SELL,
we were able to enumerate all cases in the proof of a permutation
lemma, and to decide which of these cases work~\cite{NigamTPLP13}.
The cases in such proofs are the different ways one rule can be applied
over another.
Using context constraints for each SELL rule, the possible derivations
of a formula is computed by a logic program that finds all models that
satisfy a set of constraints.
Once all possible derivations of a rule over another is found,
another logic program computes whether provability of some premises
imply provability of others.
The check is sound, but not complete.
\end{paragraph}

\begin{paragraph}{Implementations}
The goal of this work was to provide automated ways to check for some
meta-properties of sequent calculus systems, so it is only natural
that we have implemented the solutions.
Initial expansion and cut elimination for systems encoded in SELL are
implemented in the tool
Tatu\footnote{\url{http://tatu.gisellereis.com/}}.
Permutation lemma for systems encoded in SELL is implemented in the
tool
Quati\footnote{\url{http://quati.gisellereis.com/}}.
Both tools require the user to input only the encoding, and all checks
are done with the click of a button.
Tatu also features a nice interface that shows the encoded sequent
calculus rules and cases for permutation lemmas in
\LaTeX~\cite{NigamIJCAR2014}.
\end{paragraph}

\begin{paragraph}{Extensions}
This framework was adapted to the linear nested sequent setting, and
it was shown that it can more naturally capture a range of systems
with context side conditions~\cite{lns}.
\end{paragraph}

The works on using SELL as a logical
frameworks~\cite{NigamJLC16,NigamTPLP13} were successful in finding
decidable conditions on encodings that translate into meta-properties
of the encoded systems.
These conditions can be checked completely automatically, as witnessed
by their implementations.
It is not surprising, however, that the more a meta-theory proof
deviates from the ``standard'' procedure, the fewer cases can be checked
automatically.
But probably the biggest issue with using SELL to encode systems is
coming up with the encoding in the first place.
It turns out that, for encoding one system, different subexponential
configurations can be used.
Each choice might influence on which meta-properties can and cannot be
proved (even if the encoded system is correct), and figuring this out
requires patience and experience.

\section{Formalizations in Proof Assistants}

Proof assistants are incredibly expressive and powerful tools for
programming proofs.
Specifications are written usually in a relational or functional
fashion, and proofs about such specifications are written as proof
scripts.
Those scripts basically describe the proof steps, and each step is
validated by a proof checker.
If the proof assistant is able to execute the proof, and its
implementation is trustworthy, then this is a strong guarantee that
the proof is correct.

One of the issues when developing proofs of meta-properties by hand is
the sheer complexity and number of cases.
By implementing these proofs in proof assistants, the computer will
not let us skip cases or overlook details.

There are several works that implement different calculi and proofs of
meta-properties in proof assistants.
We mention a few, though this is far from an exhaustive list.
Dawson and Gor\'{e} proposed a generic method for formalizing sequent
calculi in Isabelle/HOL, and implemented meta-properties parametrized
by a set of rules~\cite{Dawson10}. 
This implementation was used to prove cut elimination of the
provability logic GLS$_V$.
Recently, D'Abrera, Dawson, and Gor\'{e} reimplemented this framework
in Coq and used it to implement and prove meta-properties of a linear
nested sequent calculus~\cite{dabrera21}.
Tews formalized a proof of cut elimination for coalgebraic logics in
Coq, which uncovered a few mistakes in the original pen and paper
proof~\cite{Tews13}.
Graham-Lengrand formalized in Coq completeness of focusing, a proof
search discipline for linear logic~\cite{lengrand14}.
Urban and Zhu formalized strong normalization of cut elimination for
classical logic in Isabelle/HOL~\cite{Urban08}.

The fact that each of these works is a publication (or collection of
publications) itself is evidence that formalizing meta-theory is far
from trivial work and cannot be done as a matter of fact.
Even though one would think that specifying sequent calculi on a
functional or relational language would be ``natural'', there is a lot
of room for design choices that influence how proofs can be
implemented.
I have been involved on the formalization of linear logic and its
meta-theory in two proof assistants: Abella~\cite{ChaudhuriTCS17} and
Coq~\cite{OlarteLSFA17}.
The Abella formalization includes various fragments of linear logic
and different calculi, and the meta-theorems proved were identity
expansion, cut elimination, and invertibility lemmas when needed.
The Coq formalization was done for first order classical LL, and
includes the meta-theorems of cut elimination, focusing, and
structural properties when needed.
The choice of linear logic is strategic: this logic's context is not a
set, so we cannot leverage the context of the proof assistant to store
the object logic's formulas.
Many substructural logics would have similar restrictions, so a
solution for linear logic can probably be leveraged for other logics
as well.
We discuss now the main challenges and insights of those
formalizations.

\subsection{Specification of Contexts}

As mentioned, we could not use Abella or Coq's set-based context to
store linear logic formulas, since these need to be stored in a
multiset.
As a result, the context needs to be explicit in the specification of
the sequent calculus.
Proof assistants typically have a really good support for lists, so
one choice would be to encode contexts as lists.
In this case, we would need to show that exchange is height-preserving
admissible, and use this lemma each time we need to apply a rule on a
formula at the ``wrong position''.
To avoid this extra bureaucracy, and to have specifications and proofs
that resemble more what is done on pen and paper, we need to use a
multiset library.

For the Abella development we implemented this library from scratch.
Multi-sets can be specified in a number of ways, with different
operations as its basic constructor.
The implementation, operations, and lemmas proved about multisets
highly influence the amount of bureaucracy in the meta-theory proofs.
Our implementation uses an \texttt{adj} operation on an element and a
multiset to add the element to the multiset (akin to list cons).
Using \texttt{adj} we could define \texttt{perm} (equality up to
permutation) and \texttt{merge} (multiset union).
It is possible that our implementation can be further simplified, but
we run the risk of over-fitting it for the linear logic case.
%

Coq has a multiset library where multisets are implemented as bags of
type \texttt{A -> nat}.
It turns out that this implementation of multisets complicates
reasoning, so a multiset library was again implemented from scratch.
In this case, a multiset is defined as a list, and multiset operations
including equality are defined inside a Coq module.

\subsection{Handling Binders}

Linear logic quantifiers are \emph{binders}. 
Encoding object level binders has been the topic of extensive research
during the last
decades~\cite{bindersBenchmark,bindersChallenge1,bindersChallenge2,bindersComparison,two-level}.

In the Coq formalization, LL quantifiers were encoded using the
technique of \emph{parametric} HOAS~\cite{phoas}, so substitution and
freshness conditions come for free. 
However, substitution lemmas and structural preservation under
substitutions must be assumed as axioms.

In the Abella formalization, the HOAS technique was also employed, but
object level binders can be modelled using Abella's nominal quantifier
$\nabla$.

To have a more smooth treatment for binders, the work
in~\cite{OlarteLSFA17} was adapted to use the Hybrid
framework~\cite{hybrid,OlarteMSCS21}.
This has allowed the formalization of a completely internal proof of
cut elimination for focused linear logic.
Moreover, the encoded linear logic was used as a meta-language for
encoding and proving properties of other systems, effectively
providing formal proofs for the theorems in the previously discussed
work~\cite{llinda}.

\subsection{Proof Development}

The proofs in both Abella and Coq were carried out faithfully to what
is usually done with pen and paper.
The technique used in Abella was structural induction on proof trees
and formulas, wheread Coq's proofs are done mostly on induction on the
derivation's height (which is explicit on the specification).

As expected, a lot of details and non-cases that are usually dismissed
on paper need to be properly discharged on a proof assistant.
As a result, the proof includes some bureaucracy and requires a lot of
patience and attention to detail.
Using Coq's powerful tactics language Ltac, the work~\cite{OlarteLSFA17}
included the implementation of tailored tactics to solve recurring
proof goals.
The amount of work pays off though, since a formalized proof is a more
trustworthy proof.

\medskip

Ultimately we are looking for a ``canonical'' way for specifying proof
systems on proof assistants, such that meta-property proofs can be
done more easily, hopefully using pieces from other proofs (much like
it is done on pen and paper).
I believe it is safe to say we are not there yet.
In the current state of the art, formalizations of meta-theory on
proof assistants serve more to increase trust than to facilitate
meta-reasoning.
But at each new formalization we learn new techniques, and maybe in
the future we could use those to implement a libraries that actually
facilitate meta-reasoning.

\section{User-friendly Implementations}

The two previous techniques for meta-reasoning aimed at complete
automation, or complete trust.
But we do not have to restrict ourselves to extremes.
Considering the operations that need to be done during meta-reasoning,
there are a number of them which could be delegated to a computer,
without requiring a lot of effort or expertise.
Some examples are: checking if two sequents are the same; enumerating
all possible ways two sequents can be unified; enumerating all
possible ways a rule can be applied; checking if a structure is
smaller than another one; computing and applying substitutions; etc.
You might even have implemented a couple of those in some context.
Most of these tasks have well-established, sound and complete,
algorithms, so they could be implemented in an easy-to-use tool to
help logicians with boring meta-reasoning proofs.
I have been involved in at least two such projects:
GAPT~\cite{EbnerIJCAR16} and Sequoia~\cite{Sequoia}.

\subsection{GAPT}

GAPT\footnote{\url{https://www.logic.at/gapt/}} stands for General
Architecture for Proof Theory, and it is a software for investigating,
implementing, and visualizing proof transformations.
This system grew from years of proof theorists collaborating with the
implementation of the algorithms they were studying at the time.
Coupled with an organized and systematic software engineering, it was
possible to build a common basis for all the different proof systems
and transformations.

Terms, formulas, and sequents are among GAPT's built-in datatypes.
In addition to that, it contains implementations of the sequent
calculus for classical logic LK, Gentzen's natural deduction system,
and resolution.
GAPT is able to import proofs from various automated theorem provers,
but it also includes its own prover.
Users have the option of building their own LK proof using GAPT's
tactic language: gaptic.
The imported proofs can be manipulated using GAPT's API, and
visualized using the GUI (which also provides limited manipulation
options).

Among the proof transformations implemented in GAPT, we highlight some
well-known ones:

\begin{itemize}
  \item Gentzen's cut elimination;

  \item skolemization: removing positive occurrences of $\forall$ and
  negative occurrences of $\exists$;

  \item interpolation: given a proof of $\Gamma_1, \Gamma_2 \vdash
  \Delta_1, \Delta_2$, find $I$ such that $\Gamma_1 \vdash \Delta_1,
  I$ and $\Gamma_2, I \vdash \Delta_2$ are provable and $I$ contains
  only predicates that appear in both partitions;

  \item translation from LK to natural deduction;
\end{itemize}

Extending GAPT with another calculus or proof transformation requires
one to delve into its development, understanding the code and its
modules.
However, most of the infrastructure is already there, and there are
many implementations to take inspiration from.
GAPT was not built for meta-reasoning specifically, but given the
amount of proof transformations involved in meta-theory, it is a good
platform for experimenting with how a calculus behaves under these
transformations.
It is not far fetched to think of implementing, for example, the
enumeration of ways a rule can be applied on a sequent, given that the
notion of sequent and rules already exist.

\subsection{Sequoia}

Sequoia\footnote{\url{https://logic.qatar.cmu.edu/sequoia/}} is a
web-based tool for helping with meta-theory of sequent calculi.
It was built to be as user-friendly as possible.
Users input their calculi in \LaTeX, and are able to build proofs in
it by clicking on a sequent and on the rule to be applied.
If the rule can be applied in more than one way, the system computes
all possibilities and prompts the user to choose one option.
The proof tree is rendered in \LaTeX\ on the fly, and users can undo
steps if they need to backtrack.

When it comes to meta-reasoning, sequoia helps by listing all cases it
can automatically deduce if the proof follows the ``usual''
strategy.
For example, for identity expansion sequoia will try to build small
derivations of size at most 2 for each connective, and check that the
open premises could be closed with identity on smaller formulas.
If it is able to do so, it will print the small derivation in \LaTeX\
so that the user can check for themselves.
The meta-properties sequoia supports are: identity expansion,
weakening admissibility, permutability of rules, cut elimination, and
rule invertibility.

\section{Conclusion}

I have discussed three different methods for reasoning about
meta-properties of (mostly sequent calculus) proof systems.
This is a biased view from my own experience, and should not be taken
as the only ways to do meta-reasoning on the computer.
Each method has its own advantages and disadvantages, which makes them
incomparable.

The solution using (subexponential) linear logic as a framework has
the great advantage that all checks can be completely automated, and a
``yes'' means the property holds, once and for all.
However, if the meta-theory proof follows a more esoteric strategy, it
is unlikely that this method will work.
A ``no'' is actually ``don't know'', and the logician is left on their
own to check the proof by hand.
Another challenge with this technique is the fact that the user needs
to be quite familiar with linear logic to be able to come up with a
reasonable encoding.

Formalizations in proof assistants have the advantage that one needs
to know simply the specification and scripting languages.
A familiarity with available libraries and techniques is helpful, but
not crucial.
Similar to the SELL solution, if the system checks the proof, then one
can be sure the property holds.
The flexibility in writing proofs allows for the more esoteric proofs
to be checked, but the cost of this is less automation.
In contrast to the previous approach, the check cannot be done with
the click of a button, and requires the logician to go through the
long process of implementing the proof and all its cases and details.

The last solution is the least ambitious one, but probably the most
realizable in the short term.
We can use the computer to aid in parts of meta-reasoning, while
leaving the most complicated cases for us to think about.
We have seen how a well-designed framework can serve as a platform to
test various proof transformations, and how a user-friendly system can
allow some easy cases to be computed automatically.

\smallskip

In the end, all solutions have their limitations, and we are still far
from a situation where meta-theory proofs can be developed in a more
trustworthy fashion.
There is still a lot of work to be done into making each of these
approaches easier to use and broader.
We are slowly making progress.

\bibliographystyle{eptcs}
\bibliography{references}

\end{document}